\documentstyle[aps,prl,preprint]{revtex}
\begin{document}
\draft
\preprint{BARI-TH 375}
\date{April 2000}
\title{
Clustering data by inhomogeneous chaotic map lattices}
\author{L. Angelini, F. De Carlo, C. Marangi, M. Pellicoro, S. Stramaglia
}
\address{
Dipartimento Interateneo di Fisica\\
Istituto Nazionale di Fisica Nucleare, Sezione di Bari\\
via Amendola 173, 70126 Bari, Italy}
\maketitle
\begin{abstract}
A new approach to clustering, based on the physical properties of inhomogeneous coupled chaotic maps,
is presented. A chaotic map is assigned to each data-point and short range couplings are
introduced. The stationary regime of the system corresponds to a macroscopic attractor independent of
the initial conditions. The mutual information between pairs of maps serves to partition the data set in
clusters, without prior assumptions about the structure of the underlying
distribution of the data. Experiments on simulated and real data sets show the effectiveness of the
proposed algorithm.
\end{abstract}


The clustering problem consists of partitioning $N$ given points into K groups 
(clusters) so that two points belonging to the same group are, in some sense, more 
similar than two that belong to different ones \cite{fu}; it has applications in 
several fields such as pattern recognition \cite{pr}, learning \cite{lea} and astrophysics
\cite{astro}.
Data points are specified either in terms of their coordinates in a D-dimensional
space or, alternatively, by means of an $N\times N$ "distance matrix" whose 
elements measure the dissimilarity of pairs of data points. This problem is 
inherently ill-posed, i.e. any data set can be clustered in drastically different 
ways, with no clear criterion to prefer one clustering over another. The most 
important sources of ambiguity are the choice of the number of clusters and the fact
that a satisfactory clustering of data depends on the desired {\it resolution}.

When prior knowledge of the clusters' structure is available (e.g., each cluster can
be represented by a multivariate Gaussian distribution), parametric approaches can be used
so that prior information is incorporated in a global criterion, thus converting clustering
onto an optimization problem. Examples of parametric clustering algorithms 
are variance minimization \cite{rose} and
maximum likelihood \cite{barkai}. In many cases of interest, however, there is no a priori
knowledge about the data structure. Then it is more natural to adopt nonparametric approaches,
which make fewer assumptions about the model and therefore are suitable to handle a 
wider variety of clustering problems. Usually these methods employ a local criterion to build
clusters by utilizing local structure of the data (e.g., by identifying high-density regions
in the data space) \cite{fu}.

A very interesting nonparametric approach for clustering, based on the physical properties
of an inhomogeneous Potts model, has been recently proposed \cite{blatt} and has proven
to perform better than other nonparametric methods. The central feature of this method, called 
{\it superparamagnetic clustering} (SPC),
is to change the similarity index of the problem from the interpoint distance to the spin-spin
correlation function of the statistical model; the temperature of the Potts model controls 
the resolution at which data are clustered.

In the present work we propose a new nonparametric method based on the physical properties of
inhomogeneous coupled chaotic maps. We assign a map to each data point and introduce
couplings, between pairs of maps, whose strength is a decreasing function of their distance.
The mutual information between pairs of maps, in the stationary regime, is then used as the 
similarity index for clustering the data set.

Systems of diffusively coupled chaotic maps, living on regular lattices, have been extensively 
studied; for large coupling strength they exhibit non-trivial collective behavior, i.e. 
long-range order emerging out of local chaos \cite{chate}. Globally coupled chaotic maps, a 
mean field extension of coupled map lattices, have also been considered and their rich variety
of behaviors has been outlined \cite{kaneko}; it has been shown that mutual synchronization of
chaotic oscillations is a robust property displayed by globally coupled maps and clusters of synchronized maps
appear in the stationary regime. In a recent paper \cite{manrubia} randomly coupled maps were studied and
the formation of dynamical clusters of {\it almost} synchronized maps was observed. Here we associate a system
of chaotic maps to a given data set so that the architecture of the network bias the formation of clusters
of {\it almost} synchronized maps to correspond to high density regions in the data set.
Let us introduce coupled chaotic maps on finite size
inhomogeneous lattices. Given a set of $N$ points $\{ {\bf r}_i\}$ in a D-dimensional
space, we assign a real variable $x_i \in [-1,1]$ to each point and define pair-interactions
$J_{ij}=\exp{(-[{\bf r}_i-{\bf r}_j]^2 /2a^2)}$, where $a$ is the local length scale. The time
evolution of the system is given by:
\begin{equation}
x_i (t+1) = {1\over C_i} \sum_{j\ne i} J_{ij} f\left(x_j(t)\right),
\label{eq:1}
\end{equation}
where $C_i =\sum_{j\ne i} J_{ij}$, and we choose the logistic map $f(x)=1-2x^2$. We note that the equivalent dynamics to
(\ref{eq:1}) in terms of variables $y_i (t)=f(x_i(t))$ is 
\begin{equation}
y_i(t+1)=f\left({1\over C_i}\sum_{j\ne i} J_{ij} y_j(t)\right);
\label{eq:2}
\end{equation} 
this form
may be more familiar for researchers in neural networks, $f$ playing the role of a nonmonotonic transfer function
\cite{carop}.
A detailed 
analysis of the behavior of this class of models will be given elsewhere \cite{prepar}; here we only 
describe some properties which will be useful for clustering purposes.

The stationary regime of Eqs.(\ref{eq:1}) corresponds to a macroscopic attractor which is independent
of the initial conditions. To study the correlation properties of the system, we consider the mutual 
information $I_{ij}$, between variables $x_i$ and $x_j$, whose definition is the following \cite{wiggins}.
If the state of element $i$ is $x_i(t)>0$ then it will be assigned a value 1, 
otherwise it will be assigned 0: this generates a sequence of bits, in a certain time interval, which
allows the calculation of the Boltzmann entropy $H_i$ for the $i$th map. In a similar way the joint entropy
$H_{ij}$ is calculated for each pair of maps and finally the mutual information is given by
$I_{ij}=H_i +H_j -H_{ij}$. The mutual information is a good measure of correlations \cite{sole} and it
is practically precision independent, due to the rough coarse graining of the dynamics. If maps $i$ and $j$
evolve independently then $I_{ij}=0$; if the two maps are exactly synchronized then the mutual information
achieves its maximum value, in the present case $ln\;2$, due to our choice of $f$.

Let us now describe our simulations of large systems (up to $N=100,000$) randomly generated with 
uniform density $\rho$ in dimension D.
The average mutual information between two maps at distance $r$ obeys the following scaling form:
\begin{equation}
I(r,a,\rho,N)=I_D \left({r\over a}\right),
\label{eq:3}
\end{equation}
where $I_D$ is a scaling function which depends on $D$ but it is independent of $N$ and $\rho$, provided that $a$
is much less than the linear size of the system. In Fig.1 we show the scaling function for $D=2$, $3$ and $4$; we see that
full synchronization is never achieved even for very close pairs of maps, indeed for $r$ close to zero $I_D$ 
is less than $ln\;2$. At 
large distances $I_D$ tends to a non-vanishing value, i.e. the system is characterized by long range
correlations. Moreover the asymptotic value $I_D (\infty )$ increases with the dimension $D$; in the limit $D\to\infty$
the system becomes a mean-field model and it can be expected that in this limit the system fully synchronizes.
Now we give the definition of $k$-nearest neighboring sites for our lattices: sites $i$ and $j$ are nearest neighbors
if and only if $j$ is one of the $k$ nearest points of $i$ and $i$ is one of the $k$ nearest points of $j$. The 
typical distance between two nearest neighbors obviously depends on the density $\rho$. Due to the scaling law
(\ref{eq:3}), it follows that, at fixed $a$, the typical amount of mutual information between nearest neighboring maps 
depends monotonically on the density $\rho$. 
Let us now turn to consider a real data set, made of regions with different densities: we find 
that the mutual information between two neighboring maps, in this case, depends on the local density around
the pair. In particular it is small in low-density regions. Our algorithm employs the contextual character of the
mutual information for clustering the data set.

Now we describe our method. The value of $a$ is fixed as the average distance of $k$-nearest neighbors pairs 
of points in the whole system (our results are quite insensitive to the particular value of $k$). We remark that 
everything
done so far can be easily implemented in the case when instead of providing the $\{ {\bf r}\}$ for all data we have an 
$N\times N$ "distance matrix". For the sake of computational convenience, we keep only interactions of a map
with a limited number of maps, those whose distance is less than $3a$, and set all other $J_{ij}$ to zero. Starting
from a random initial configuration of $\{x\}$, Eqs.(\ref{eq:1}) are iterated until the system attains its stationary regime;
the mutual information is then evaluated for pairs of maps. The clusters are identified in the two following steps. 
(1) A link is set between all pairs of data points such that $I_{ij} > \theta$, where $\theta$ is a threshold. 
(2) Data clusters are identified
as the linked components of the graphs obtained in step 1. 

The value of $\theta$ controls the resolution at which the data set is clustered; 
by repeating the two steps above described for an 
increasing sequence of $\theta$ values, hierarchical clustering of the data is obtained.

The following toy problem illustrates how the proposed algorithm works. Fig. 2 contains two dense regions of $400$
and $1900$ points on a dilute background of $200$ points. In Fig.3 we show the frequency distribution of 
(a) distances between neighboring points and (b) mutual information between neighboring points. The peak around $I=0$,
in Fig. 3b, corresponds to points in the background. In Fig. 4 we show the size of the three biggest clusters, found 
by our algorithm, versus $\theta$. For $\theta <0.035$ the algorithm identifies a single big cluster of about 2400 points, 
the remaining $100$ points are distributed among $58$ clusters of size smaller than $9$. For $0.035<\theta <0.285$ two
big clusters, corresponding to the two dense regions, are identified; these clusters consist of $1913$ and $411$ points 
respectively, while the remaining $176$ points are distributed in $98$ clusters of size smaller than $9$. 
As $\theta$ increases above $0.285$ the biggest 
clusters break into smaller and smaller clusters. As can be observed from Fig.4, the stability of the largest clusters 
(existence of a plateau) is a clear indication of the optimal partition among the whole hierarchy yielded by our algorithm.
The results above described correspond to $k=20$, however values in the range $5-50$ give similar results.

Now we turn to consider a real data-set extracted from LANDSAT Thematic Mapper (TM) images. We analyze data taken 
from a satellite image of an area in the Southern of Italy consisting of $1489$ pixels each of which is represented by
six spectral bands. The ground truth was determined by means of visual interpretation of areal photos  followed by site 
visits. The area study includes seven landcover classes: (A) {\it Coniferous reafforestation}, $69$ points; (B) 
{\it Bare soil}, $85$ points; (C) {\it Urban areas}, $91$ points; (D) {\it Vineyards}, $300$ points; (E) {\it Cropland},
$316$ points; (F) {\it Pasture}, $265$ points; (G) {\it Olives groves}, $363$ points. In Fig. 5 the first two principal
components of the data-set are shown: this problem is characterized by clusters of different size and density.
In spite of these difficulties, our algorithm succeeds in resolving the data-structure, as it is shown in Fig. 6. For 
$\theta < 0.02$ our algorithm identifies a single big cluster; at $\theta = 0.02$ this cluster splits in two clusters, one 
corresponding to class A and the other corresponding to data-points of the other six classes. By successive transitions,
all the seven classes separate. Both the six clusters partition and the seven clusters one appear stable: 
prior knowledge is needed, in this case, to select the correct partition of the data set. 
In the range $0.27 <\theta < 0.35$ seven clusters, consisting of $69$, $72$, $88$, $295$,
$325$, $298$, $291$ points respectively, 
are stable; the remaining $51$ points are distributed among $18$ small clusters of size smaller than $7$.
Hence $96.6\%$ of data is classified; the purity of the classification (percentage of correctly classified points) is 
$96.2\%$.
As $\theta$ is further increased, these clusters break into smaller and smaller parts (the cluster which breaks first is 
the one corresponding to class D). It is worth mentioning that an unsupervised exploration of the 
underlying structure in a data set (like the one provided by the proposed method) make easier the design of a supervised
classifier for the same problem (see \cite{domany}).

It is clear that the proposed algorithm has similarities with SPC method, indeed both methods 
associate a physical system to data-set points and employ a physical correlation (spin-spin correlation \cite{blatt}
or mutual information) as the similarity index. We apply SPC to the LANDSAT data-set and
obtain the same hierarchical structure of data as the one from our method; the best performance corresponds
to seven clusters of $70$, $48$, $78$, $255$, $317$, $284$, $283$ points respectively: $89.6\%$
of data are classified with $96.6\%$ purity \cite{oss1}. Hence, as far as the data-set at hand is concerned, our algorithm 
classifies more points than SPC with almost the same purity.
Our algorithm has the following computational
advantage over SPC: the hierarchical structure of data
is obtained by a simple thresholding at each value of $\theta$, while SPC requires a
Monte Carlo at each value of the temperature. This reduces the computational time by orders of magnitude.
On the other hand, SPC provides a supplementary indicator, the susceptibility \cite{blatt}, 
which may be helpful to detect the optimal 
partition of the data-set.

Some remarks are in order. We have also clusterized data using the average distance between maps $|x_i-x_j|$, 
in the stationary regime, as the dissimilarity index: the results were less stable than those obtained by use of 
the mutual information. Our choice of the logistic map $f(x)=1-\alpha x^2$, with $\alpha=2$, is due to the circumstance 
that the 
corresponding invariant measure is symmetric around $0$, so that the mutual information can, in principle, achieve
its maximum value $ln\; 2$; other maps with symmetric invariant measure work as well, while choosing maps with
non symmetric measure would reduce the allowed range of values for the mutual information.    
Finally we wish to reemphasize the aspects we consider as the main advantages of our algorithm: its simplicity, the physical 
system to be simulated being described by simple deterministic equations (\ref{eq:1}), and its general 
applicability, no {\it a priori} knowledge of clusters' structure is to be assumed. Applications of our algorithm to other
real problems will be presented in a forthcoming paper \cite{prepar}.

\section*{Ackowledgements}
The authors thank G. Nardulli, for valuable discussions, and P. Blonda, G. Pasquariello, G. Satalino (IESI-CNR) 
for providing the LANDSAT
data set.

\newpage
\noindent\Large\textbf{FIG. Captions}
\normalsize
\vspace{1.0cm}
\begin{description}
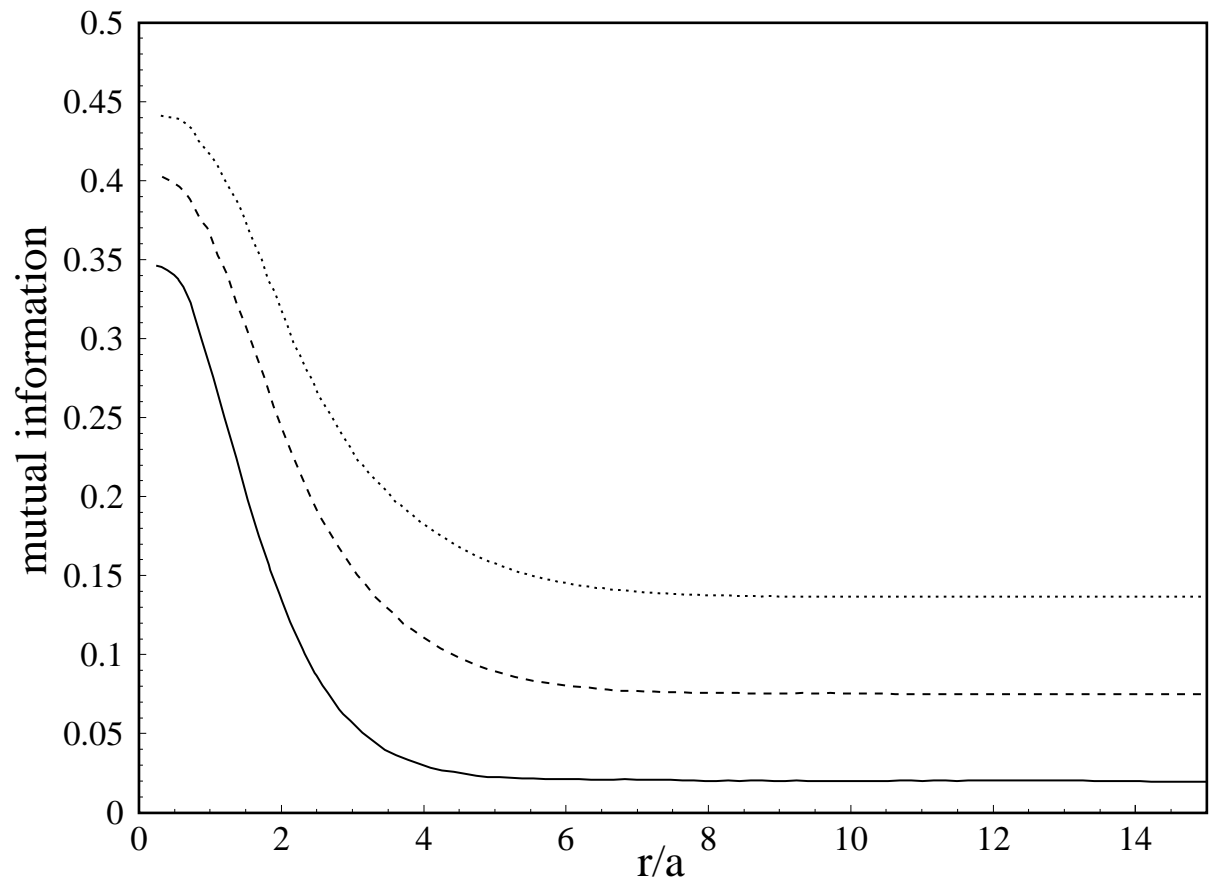
\item{FIG. 1}: Scaling function for the mutual information for $D=2$ (solid line), $D=3$ (dashed line) and $D=4$ (dotted
line).

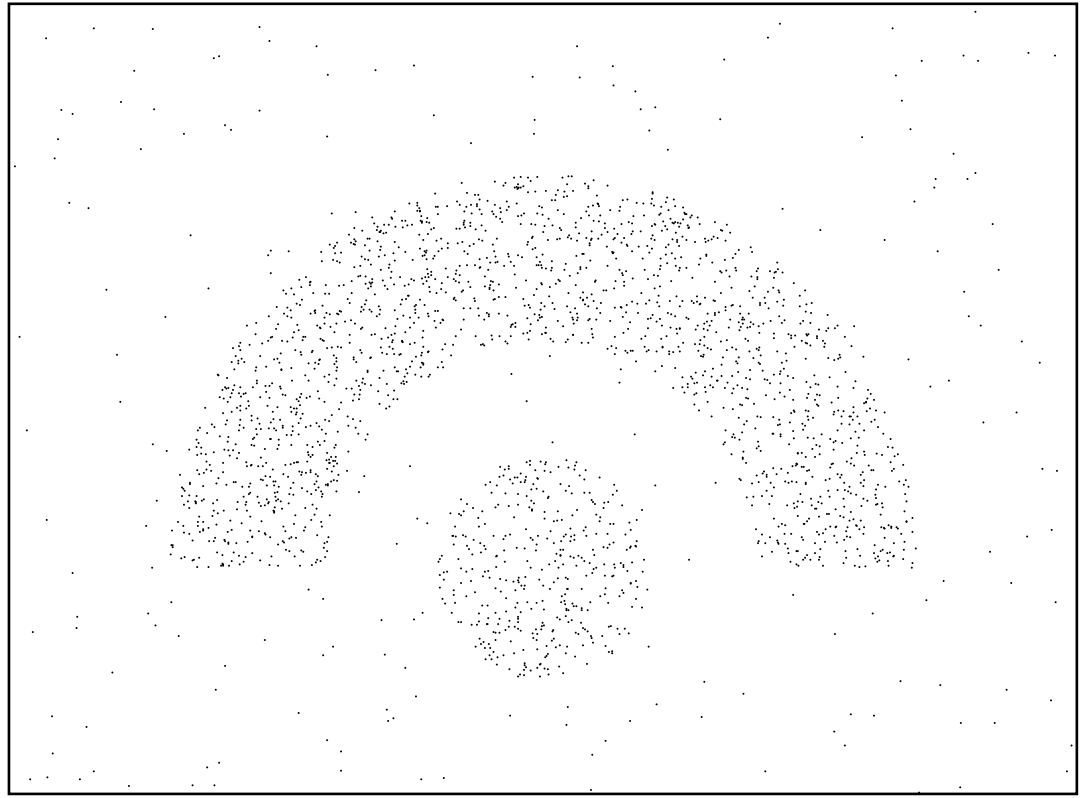
\item{FIG. 2}: Artificial data set consisting of two dense regions of $400$ and $1900$ points, in 
a dilute background of $200$ points.

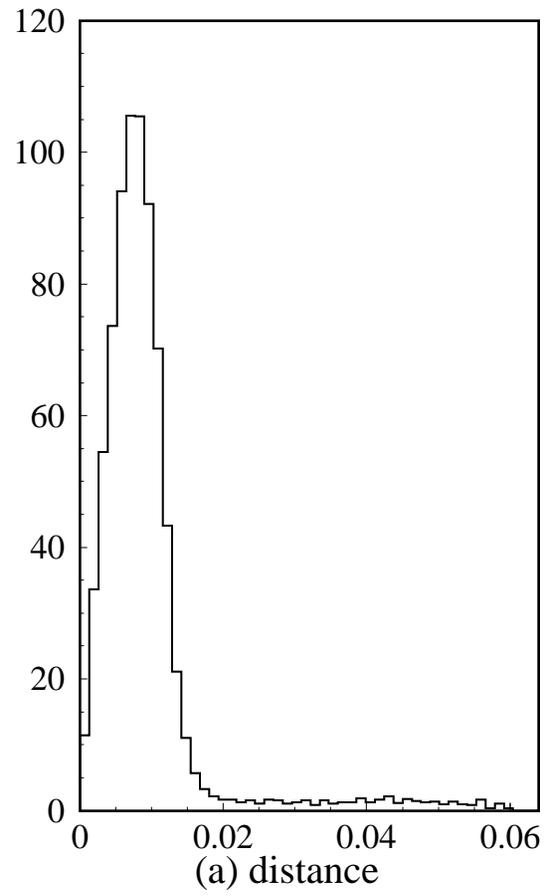
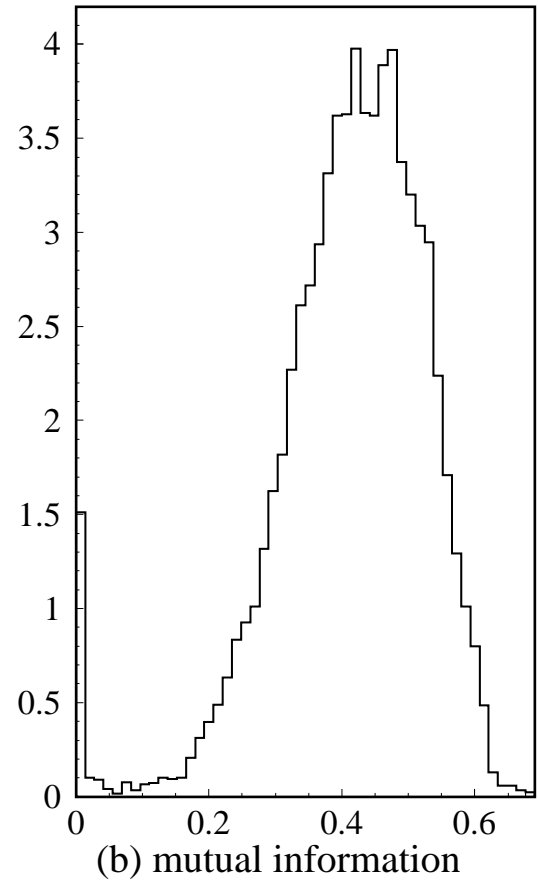
\item{FIG. 3}: Frequency distribution of (a) distances between neighboring points in Fig. 2 and (b) mutual information
of neighboring points.

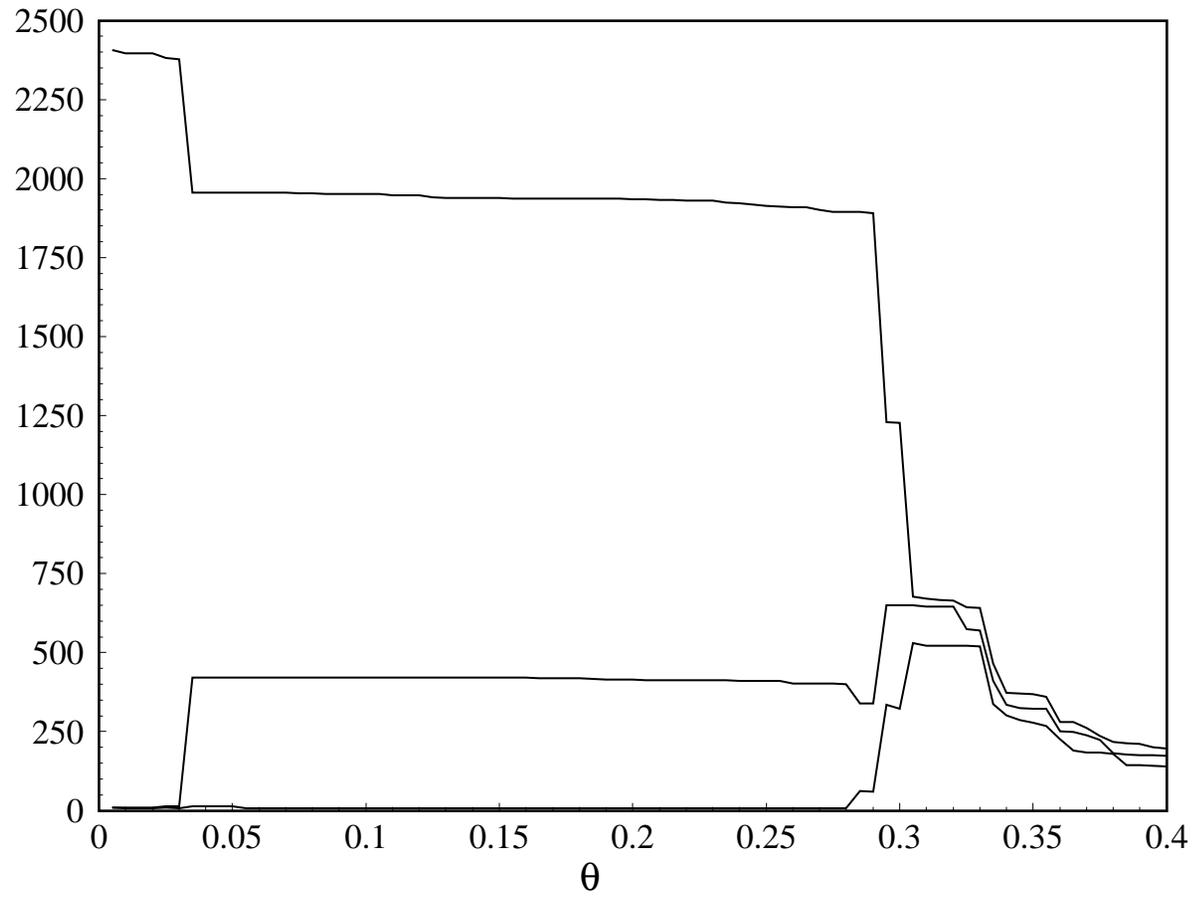
\item{FIG. 4}: Size of the three biggest clusters obtained by our algorithm, on the data-set in Fig.2, as a function
of the threshold $\theta$.

\item{FIG. 5(color)}: First two principal components of the LANDSAT data-set (see the text).
Comparing the variances along the six principal axis, it turns out that also the third and fourth principal components
are relevant to this data-set.

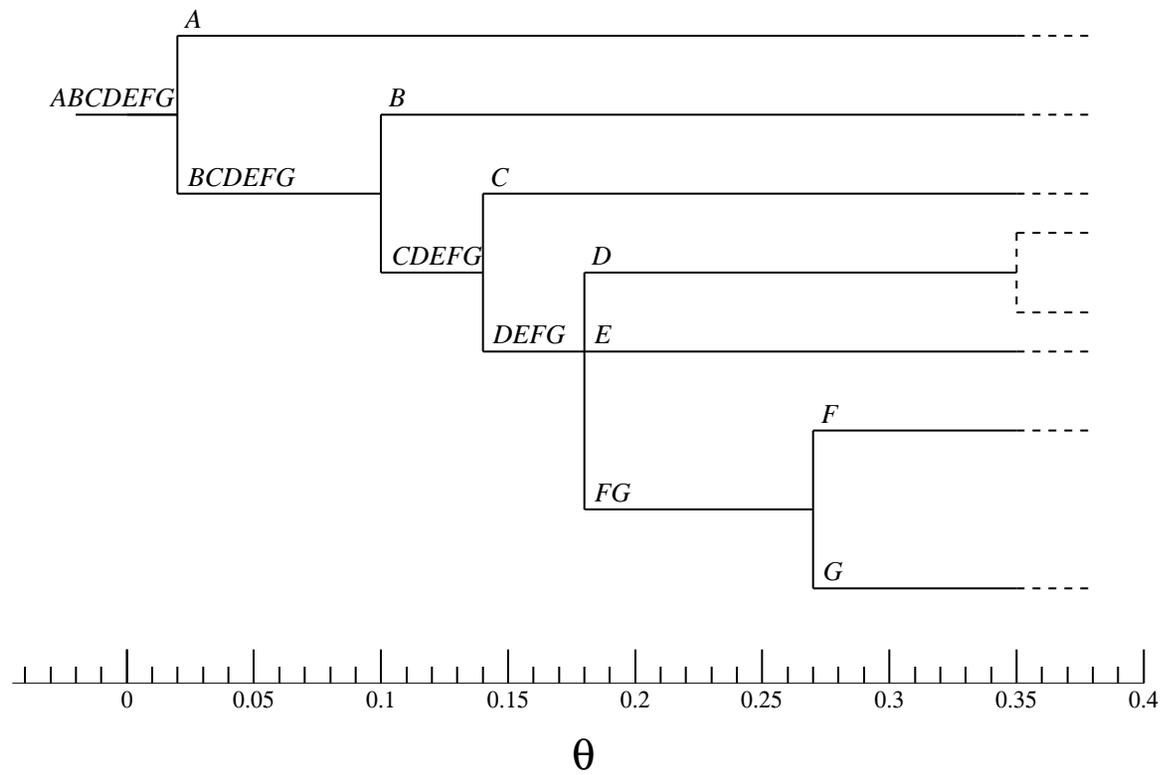
\item{FIG. 6}: Hierarchical structure of the LANDSAT data-set as it has been found by our algorithm; the $\theta$ values
at which the clusters split can be read on the axis at the bottom. This results have been obtained using $k=20$, however
values in the range $10-50$ give similar results.

\end{description}
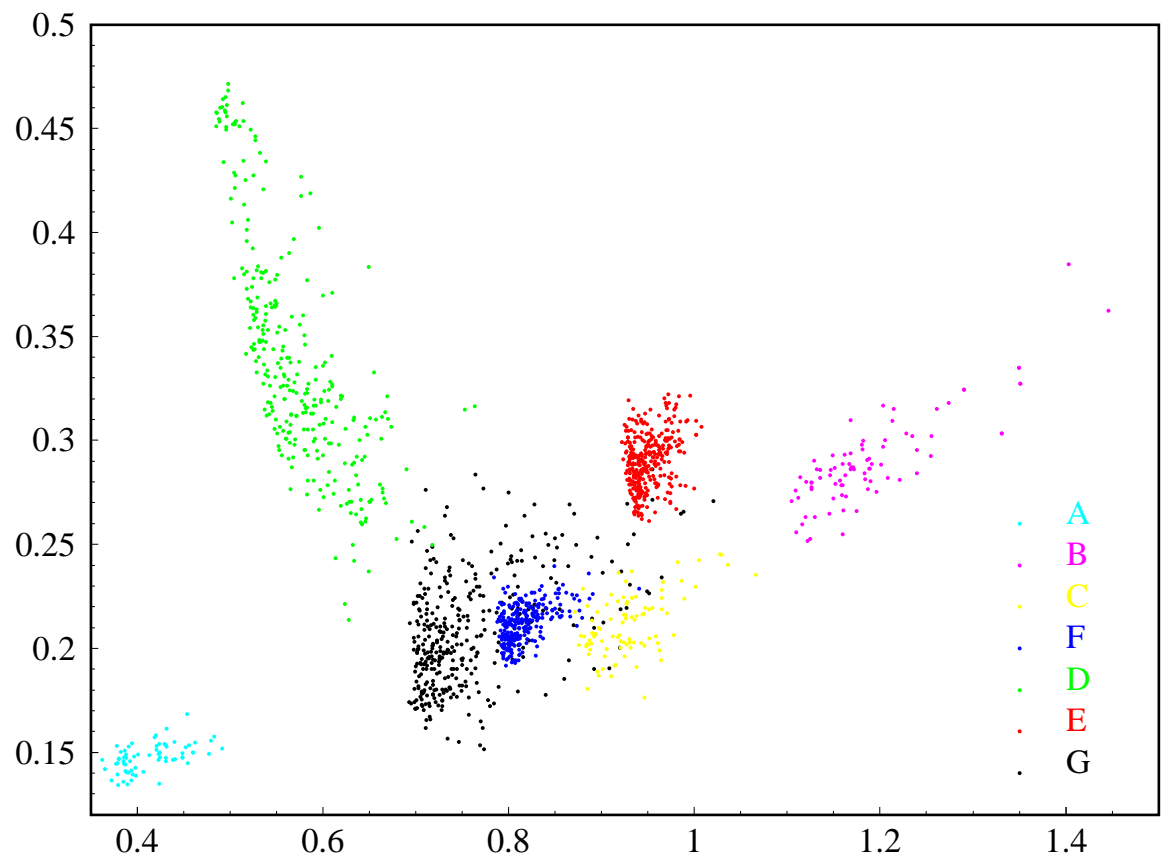
\end{document}